\documentclass[hyper,a4paper,notoc]{JHEP}

\usepackage{amsmath,amssymb}

\title{From Superpotential to Model Files for FeynArts and CalcHep/CompHep}

\author{Florian Staub \\
Institut f\"ur Theoretische Physik und Astrophysik, Universit\"at W\"urzburg,\\
D-97074  W\"urzburg, GERMANY\\
Email: \email{florian.staub@physik.uni-wuerzburg.de}
}

\abstract{
SARAH is a Mathematica package for building and studying supersymmetric models. It calculates for a given superpotential and gauge sector the full Lagrangian of a model. With the new version of SARAH it is possible to calculate automatically  all interactions for the different eigenstates and write model files for FeynArts and CompHep/CalcHep. In addition, the tadpole equations are calculated, gauge fixing terms can be given and ghost interactions are added, particles can be integrated out and non supersymmetric limits of the theory can be chosen. CP and flavor violation can easily be switched on or off. 
}

\begin{document}
\section{Introduction}
Supersymmetry is one of the most popular extensions of the standard model (SM) of particle physics \cite{susy1,susy2,susy3,susy4,susy5,susy6,susy7}. The minimal supersymmetric standard model (MSSM) is well studied and there are several tools to explore the model: For every event generator or diagram calculator exists a model file, thus the MSSM can be handled out of the box. But what happens, if you want to change the model, extend the particle content, add a new gauge group or drop some assumptions about the parameters? Before you are able to do really physics and phenomenology you have to do a lot of work before: Check if your new model is free from gauge anomalies, get the Lagrangian out of the superpotential and extract all interactions, calculate the mass matrices and diagonalize them to get your mass spectrum, derive the tadpole equations to find the minimum conditions for your potential and so on. After all, if you are content with the first impression of your model, you might want to do phenomenological studies. But before you can use one of the existing programs, you have first to write a model file for them, what is again a very time consuming work. \\
This is exactly that kind of work SARAH can now do for you. SARAH just needs the gauge structure, particle content and superpotential to produce all information about the gauge eigenstates. If you have also broken gauge symmetries and particle mixings you can easily add them. Also the gauge fixing terms can be specified. At the end you have the choice, if you want a \LaTeX{} file with all information about your model, if a model file for FeynArts \cite{FeynArts} or CalcHep/CompHep \cite{CHep} should be written or if you just want to check some things using Mathematica. This model files can of course also be used with FormCalc \cite{FormCalc} for calculating Feynman diagrams with Mathematica or micrOmegas \cite{micrOmegas} for dark matter studies. The intention by the development of SARAH was to make it very flexible: There is a big freedom for the matter and gauge sector you can handle. The work with SARAH should be easy: Every information SARAH needs are specified in an easy to modify model file. Nevertheless, SARAH is also fast: A existing model can be changed within minutes, and the needed time for doing all necessary calculations and writing a model file is normally less than 10 minutes. \\

\section{Overview}
SARAH is a package for Mathematica \footnote{Mathematica is a protected product by Wolfram Research} and was tested with versions 5.2 and 7.0. \\  
The package can be downloaded from
\begin{verbatim}
http://theorie.physik.uni-wuerzburg.de/~fnstaub/sarah.html
\end{verbatim}
and a comprehensive manual for SARAH can be found here \cite{SARAH}.

After you have extracted and saved the package, you can load SARAH in Mathematica by   
\begin{verbatim}
<<"\Package Directory\SARAH.m"
\end{verbatim}
and start the calculations for a model with
\begin{verbatim}
Start["modelname"];
\end{verbatim} 

The following steps are done automatically:
\begin{itemize}
\item The model is checked for gauge anomalies
\item The Lagrangian for the gauge and mass eigenstates is calculated
\item Ghost terms are calculated from the gauge fixing terms
\item If chosen, particles are integrated out
\item Mass matrices are derived
\item The tadpole equations are calculated
\item All tree level masses are calculated
\item A spectrum file is read in and the mixing matrices are calculated 
\end{itemize}

On modern computers, this will last only a few minutes. After the automatized calculations are finished, you can just play with the model by checking distinct vertices, look at the different mass matrices, give numerical values to the parameters and calculate the mass spectrum, or you can generate model files for FeynArts and CompHep/CalcHep for further phenomenological studies. Also a \LaTeX{} output with the particle content, mass matrices, tadpole equations and all interactions including corresponding Feynman diagrams can be written.   

\section{What Models are possible}
In principle, SARAH can handle every \(N=1\)-SUSY theory, in which the chiral superfields are arranged in \(SU(2)\) doublets and singlets. The gauge sector can be extended by abelian and also non abelian gauge groups, more particles and/or generations of particles can be added to the matter content, one or more gauge symmetry breakings can be defined, particles can be integrated out to get an effective theory or just 'deleted' to get a non supersymmetric limit of a theory like the standard model. Non canonical terms can be added to the Lagrangians and existing interactions can be redefined. \\
You have always full control over the properties of the parameters of your model: You can switch on and off flavor and CP violation independently for all parameters, define relation between the parameters, add numerical values or specify the position in a Les Houches input file. \\
In the package archive of SARAH already some models are included: The MSSM, the MSSM in CKM basis, the NMSSM, the \(\mu\nu\)SSM, the MSSM with bilinear R-parity violation, the SM, an effective MSSM with integrated out Gluino and a MSSM with additional \(U(1)\).

\section{The Model Files}
All information about a model are saved in three different files: \verb"Model.m", \verb"parameters.m" and \verb"particles.m". Only the first one is absolute necessary and contains the information about the gauge sector, particle content, superpotential and mixings. In \verb"parameters.m" you can assign properties to all parameters of your model, give numerically values and define the \LaTeX{} name for each parameter. In \verb"particles.m" additional information about the particles are given, which might be needed for an appropriate output: R-parity, mass, width, PDG, \LaTeX{} name and output name. \\
All three files are written in an intuitive way and can easily and quickly be changed. This, we want to demonstrate at an example in the appendix.

\section{The Output}
With
\begin{verbatim}
ModelOutput[EWSB,WriteTeX->True, WriteCHep->True, WriteFeynArts->True];
\end{verbatim}
it is possible to create model files for FeynArts and CalcHep/CompHep as well as a \LaTeX{} file with all information about the model for given eigenstates.
\subsection{\LaTeX}
In the generated \LaTeX{} files are all information about a model for one set of eigenstates included: Particle content, mixing matrices, tadpole equations and all interactions. For the interactions are also automatically the corresponding Feynman diagrams are drawn by using the \LaTeX{} package \verb"FeynMF" of Thorsten Ohl \cite{ohl}. For generating the pdf file from the different \LaTeX{} files and all Feynman diagrams a batch file is written automatically by SARAH for Linux and for Windows. 

\subsection{CalcHep/CompHep}
There have to be taken some difficulties into account when writing a model file for CompHep or CalcHep: The color structure is in implicit, therefore a four point interaction of four colored particles can't be handled. This was solved by the authors of CalcHep/CompHep by adding auxiliary fields and splitting such interactions in two parts. SARAH does the same splitting by taking the results from the F- and D-Terms. This has also the nice advantage that not all possible combinations of 4 scalar have to be written separately and leads to an increasing of the speed writing the model file. \\
In addition, CalcHep/CompHep works with three different kind of ghosts: Fadeev Popov ghost, Goldstone ghosts and tensor ghosts. The first two are supported by SARAH and the vertices are calculated dynamically, i.e. the model files produced by SARAH support both Feynman gauge as well as unitarity gauge. The tensor ghost of the gluino and its one interaction is hard coded.\\
CalcHep/CompHep doesn't support complex parameters. Therefore SARAH has an option to split each vertex in a real and imaginary part in the Lagrange file of CalcHep/ CompHep in order to enable also theories with CP violation. \\

\subsection{FeynArts}
Generating the FeynArts model file is more straight forward than the files for CalcHep/ CompHep. But in addition to the normal model file for FeynArts SARAH writes a second file including additional information about the model: All defined dependences, the numerical values for the parameters and masses if they are available and special abbreviations to speed up the calculations with FormCalc.

\section{Checks}
We have checked the model files generated with SARAH  for the MSSM against the existing files of FeynArts and CalcHep both at the vertex level and at the process level. This means, we have compared the value of each vertex for a given set of parameters and all possible combinations of generations (more than 15000), and we have also calculated several processes with the old and new model files. In addition we have calculated the relic density with micrOmegas for two different set of parameters. \\
A recent version of SARAH was also used for \cite{staub2} and the results were cross checked in several ways. Also one of the authors of \cite{mnSSM2} checked the results of SARAH for the $\mu\nu$ SSM.   

\section*{Acknowledgments}
This work is supported by the DFG Graduiertenkolleg GRK-1147 and by the DAAD, project number D/07/13468. We want thank Thomas Hahn for his help concerning FeynArts and FormCalc. In addition, we thank Ritesh K. Singh for very fruitful suggestions during the development of SARAH, Stefan Liebler for his checks of SARAH and its results and Werner Porod for helpful discussions about SUSY in general. Furthermore, we are grateful to Nicole Schatz for reading the draft and her support during all the work.

\begin{appendix}    

\section{Example: From MSSM via NMSSM to $\mu\nu$SSM and more}
\subsection{The MSSM}
\subsubsection{The Model File}
First, we want to explain the different parts of the model file of the MSSM. 
\begin{enumerate}
\item The gauge sector is \(U(1)\times SU(2)\times SU(3)\) and is just defined by adding the corresponding vector superfields. 
\begin{verbatim}
Gauge[[1]]={B,  1, hypercharge, g1};
Gauge[[2]]={WB, 2, left,        g2};
Gauge[[3]]={G,  3, color,       g3};
\end{verbatim}
The different parts are: The superfield name, the dimension of the gauge group, the name of the gauge group and the name of the gauge coupling. 
\item The doublet superfields are \(q,l, H_d, H_u\) and added by 
\begin{verbatim}
Doub[[1]] = {uL,  dL,  3, q,   1/6, 1, 1};  
Doub[[2]] = {vL,  eL,  3, l,  -1/2, 1, 0};
Doub[[3]] = {Hd0, Hdm, 1, Hd, -1/2, 1, 0};
Doub[[4]] = {Hup, Hu0, 1, Hu,  1/2, 1, 0};
\end{verbatim}
The different parts are: The name of the up component, the name of the down component, the number of generations, the name of the superfield the different charges under the gauge groups. 
\item The singlets superfields are \(d, u, e\) and added by
\begin{verbatim}
Sing[[1]] = {dR, 3, d,  1/3, 0, -1};
Sing[[2]] = {uR, 3, u, -2/3, 0, -1};
Sing[[3]] = {eR, 3, e,    1, 0,  0};
\end{verbatim}
The definition is analog to the definition of the doublets. 

\item The trilinear terms  of the superpotential leading to the Yukawa interactions are
\begin{equation*}
\label{superpotential}
W_{tri} = - \hat{L} Y_e \hat{e} \hat{H}_d -  \hat{Q} Y_d \hat{d} \hat{H}_d +  \hat{Q} Y_u \hat{u} \hat{H}_u 
\end{equation*}
and given in SARAH by
\begin{verbatim}
TriW = {{q, Hu, u, Yu,1}, {q, Hd, d, Yd, -1}, {l, Hd, e, Ye, -1}};
\end{verbatim}

\item The bilinear term leading to the Higgsino mass is
\begin{equation*}
\label{superpotential}
W_{bi} = \mu \hat{H}_d \hat{H}_u 
\end{equation*}
This reads in SARAH
\begin{verbatim}
BiW = {{Hu,Hd,\[Mu],1}};
\end{verbatim}

\item The gauge fixing terms for the unbroken gauge groups are
\begin{verbatim}
GaugeFix={ {Der[VWB],  -1/(2 RXi[VWB])},
           {Der[VG],   -1/(2 RXi[VG]) }};
\end{verbatim}
This has the meaning of
\begin{equation*}
-\frac{1}{2 \xi_W}|\partial_\mu W^{\mu,i}|^2  -\frac{1}{2 \xi_g}|\partial_\mu g^{\mu,i}|^2
\end{equation*}

\item Begin with the definition for the first set of rotations
\begin{verbatim}
Rotation[[1]]={
\end{verbatim}
\begin{enumerate}
\item The name of the rotation is
\begin{verbatim}
EWSB
\end{verbatim} 

\item The vector bosons and gauginos rotate as follows
\begin{verbatim}
{ {VWB, {1,{VWm,1/Sqrt[2]},{conj[VWm],1/Sqrt[2]}},
        {2,{VWm,-\[ImaginaryI]/Sqrt[2]},
           {conj[VWm],\[ImaginaryI]/Sqrt[2]}},
        {3,{VP, Sin[ThetaW]},{VZ, Cos[ThetaW]}}},
  {VB,  {1,{VP, Cos[ThetaW]},{VZ,-Sin[ThetaW]}}},
  {fWB, {1,{fWm,1/Sqrt[2]}, {fWp,1/Sqrt[2]}}, 
        {2,{fWm,-\[ImaginaryI]/Sqrt[2]},
           {fWp,\[ImaginaryI]/Sqrt[2]}},
        {3,{fW0,1}}}                                                         
      },   
\end{verbatim}

This is the common mixing of vector bosons and gauginos after EWSB:

\begin{eqnarray*}
 W^\pm &=& \frac{1}{\sqrt{2}} W_1 \mp i W_2, \hspace{1cm} \tilde{W}^\pm = \frac{1}{\sqrt{2}} \tilde{W}_1 \mp i \tilde{W}_2 \\
 Z &=& -\sin\Theta_W B + \cos\Theta_W W_3, \hspace{1cm} 
 A = \sin\Theta_W W_3 + \cos\Theta_W B 
\end{eqnarray*}

\item The neutral components of the scalar Higgs get  Vacuum Expectation Values (VEVs)
\begin{eqnarray*}
\label{higgsvev}
 \left(\begin{array}{c}
        H_d^0 \\
	H_d^-
       \end{array}
 \right) = \left(\begin{array}{c}
\frac{1}{\sqrt{2}}( v_d + i \sigma_d + \phi_d) \\
 \phi_d^-
\end{array}
 \right) \\
\left(\begin{array}{c}
        H_u^+ \\
	H_u^0
       \end{array}
 \right) = \left(\begin{array}{c}
\phi_u^+ \\
 \frac{1}{\sqrt{2}}(v_u + i \sigma_u +  \phi_u)
\end{array}
 \right) .
\end{eqnarray*}
Here \(v_u\) and \(v_d\) are the VEVs of the Higgs. This is added to SARAH by

\begin{verbatim}
{{SHd0, {vd, 1/Sqrt[2]}, {sigmad, \[ImaginaryI]/Sqrt[2]},
                         {phid, 1/Sqrt[2]}},
 {SHu0, {vu, 1/Sqrt[2]}, {sigmau, \[ImaginaryI]/Sqrt[2]},
                         {phiu, 1/Sqrt[2]}}}
\end{verbatim}

\item The particles mix after EWSB to new mass eigenstates
\begin{verbatim}
{{{SdL, SdR}, {Sd, ZD}},
 {{SuL, SuR}, {Su, ZU}},
 {{SeL, SeR}, {Se, ZE}},
 {{phiu, phid}, {hh, ZH}},
 {{sigmau, sigmad}, {Ah, ZA}},
 {{conj[SHup],SHdm},{Hpm,ZP}},
 {{fB, fW0, FHd0, FHu0}, {L0, ZN}}, 
 {{{fWm, FHdm}, {fWp, FHup}}, {{Lm,Um}, {Lp,Up}}}} 
\end{verbatim}
The different parts are
\begin{itemize}
\item Mixing of down squarks:
\begin{eqnarray*}
\tilde{d}_L^I &=& S_D^{Ii*} \tilde{D}_{m,i}, \hspace{1cm}\tilde{d}_R^I = S_D^{(I+3)i} \tilde{D}^*_{m,i} .
\end{eqnarray*}
\item Mixing of up squarks 
\begin{eqnarray*}
\tilde{u}_L^I &=& S_U^{Ii} \tilde{U}_{m,i}, \hspace{1cm} \tilde{u}_R^I = S_U^{(I+3)i*} \tilde{U}^*_{m,i},
\end{eqnarray*}
\item Mixing of sleptons
\begin{eqnarray*}
\tilde{e}^I_L &=& S_E^{Ii*} \tilde{E}_{m,i}, \hspace{1cm} \tilde{e}^I_R = S_E^{(I+3)i} \tilde{E}^*_{m,i} .
\end{eqnarray*}
\item Mixing of neutral, scalar Higgs
\begin{equation*}
\label{rothiggs1}
 \left(\begin{array}{c}
        h_1 \\
	h_2
       \end{array}
 \right) = Z^H \left(\begin{array}{c}
        \phi_d \\
	\phi_u
       \end{array}
 \right)
\end{equation*}
\item Mixing of neutral, pseudoscalar Higgs
\begin{equation*}
\label{rothiggs2}
 \left(\begin{array}{c}
        A^h_1 \\
	A^h_2
       \end{array}
 \right) = Z^A \left(\begin{array}{c}
        \sigma_d \\
	\sigma_u
       \end{array}
 \right)
\end{equation*}
\item Mixing of charged Higgs
\begin{equation*}
 \left(\begin{array}{c}
        H^-_1 \\
	H^-_2
       \end{array}
 \right) = Z^\pm \left(\begin{array}{c}
        \phi_d^- \\
	\phi_u^-
       \end{array}
 \right)
\end{equation*}
With \(\phi_u^- = (\phi_u^+)^* \).
\item Mixing of neutralinos: \\
\((\tilde{B}, \tilde{W}, \tilde{H}_d^0, \tilde{H}_u^0)\) mix due to the matrix \(Z^N\) to \(\lambda^0_i\) which form Majorana fermions
\begin{equation*}
\chi_i = \left( \begin{array}{c} \lambda_i^0 \\
\bar{\lambda}_i^0
\end{array}
\right),
\end{equation*}.   
\item The last entry is the mixing of the charginos: The gauge eigenstates \(\tilde{W}^-\) (\verb"fWm") and \(\tilde{H}_d^-\) (\verb"FHdm") are mixed to the negative charged mass eigenstates \(\lambda^-\) (\verb"Lm"), while \(\tilde{W}^+\) (\verb"fWp") and \(\tilde{H}_u^+\) (\verb"FHup") form the new eigenstates \(\lambda^+\) (\verb"Lp"). The new and old eigenstates are connected by the mixing matrices \(U^-\) (\verb"Um") and \(U^+\) (\verb"Up").
\begin{equation*}
\left( \begin{array}{c} \lambda^-_1 \\ \lambda^-_2 \end{array} \right) = U^- \left( \begin{array}{c} \tilde{W}^- \\ \tilde{H}_d^- \end{array} \right), \hspace{1cm}
\left( \begin{array}{c} \lambda^+_1 \\ \lambda^+_2 \end{array} \right) = U^+ \left( \begin{array}{c} \tilde{W}^+ \\ \tilde{H}_u^+ \end{array} \right) 
\end{equation*} 

\end{itemize}

\item The new gauge fixing terms are
\begin{verbatim}
{{Der[VP],    - 1/(2 RXi[VP])},	
 {Der[VWm]+\[ImaginaryI] Mass[VWm] RXi[VWm] Hpm[{1}], 
              - 1/(RXi[VWm])},
 {Der[VZ] + Mass[VZ] RXi[VZ] Ah[{1}],  
              - 1/(2 RXi[VZ])},
 {Der[VG],    - 1/(2 RXi[VG])}}
\end{verbatim}

\end{enumerate}
\item No particles should be integrated out or deleted
\begin{verbatim}
IntegrateOut={};
DeleteParticles={};
\end{verbatim}

\item The Dirac spinors are the following
\begin{verbatim}
dirac[[1]] = {Fd,  FdL, FdR};
dirac[[2]] = {Fe,  FeL, FeR};
dirac[[3]] = {Fu,  FuL, FuR};
dirac[[4]] = {Fv,  FvL, 0};
dirac[[5]] = {Chi, L0, conj[L0]};
dirac[[6]] = {Cha, Lm, conj[Lp]};
...
\end{verbatim}
\end{enumerate}

\subsubsection{Parameter and Particle - File}
Additional properties and information of the parameters and particles of a model are saved in the files \verb"parameters.m" and \verb"particles.m". A entry in the parameter file looks like
\begin{verbatim}
{Yu, { LaTeX -> "Y^u",
       Real -> True,
       Form -> Diagonal,
       Dependence ->  None, 
       Value -> None, 
       LesHouches -> {{{1,1}, {Yu,1,1}},
                      {{2,2}, {Yu,2,2}},
                      {{3,3}, {Yu,3,3}}}
             }}
\end{verbatim}
and contains information about the numerical value, the position in a Les Houches accord file or the dependence on other parameters. Also simplifying assumptions like only real and diagonal values can be given to a parameter and the \LaTeX{} name is defined. \\  
The particles file contains entries like 
\begin{verbatim}
{Su ,  {  RParity -> -1,
          PDG ->  {1000002,2000002,1000004,2000004,1000006,2000006},
          Width -> {0, 0, 0, 0, 0, 0},
          Mass -> Automatic,
          FeynArtsNr -> 13,   
          LaTeX -> "\\tilde{u}",
          OutputName -> "um" }},   
\end{verbatim}
and defines properties of all particles like the R-parity or the mass. 

\subsection{From MSSM to NMSSM}
\subsubsection{Changing the Model File}
We want to show at one example how easy it is, to generate model files for new SUSY models with SARAH. For this reason we want to change the existing model file of the MSSM to a model file for the NMSSM (see \cite{NMSSM} and references therein). This is done by:
\begin{enumerate}
\item A gauge singlet is added to the model. This is performed in SARAH by extending the particle content with
\begin{verbatim}
Sing[[4]] = {Sing, 1, S, 0, 0, 0}; 
\end{verbatim}
This statements means, that a superfield called \(S\) is added to the particle content, which has one generation and does not couple to the three gauge groups. 
\item There is a superpotential term, which describes the interaction between the two Higgs superfields \(\hat{H}_u\) and \(\hat{H}_d\) and the gauge singlet \(\hat{S}\). Also, there is a cubic self coupling of the singlet field \(\frac{1}{3} \kappa \hat{S}^3\). This terms are added in SARAH by
\begin{verbatim}
TriW = {..., {Hu, Hd, S, \[\Lambda],1},{S,S,S,\[\Kappa],1/3}};
\end{verbatim}
\item First, we consider a CP conserving NMSSM. This is done by declaring \(\kappa\) and \(\lambda\) as real in the parameter file:
\begin{verbatim}
{\[Kappa],     {..., Real -> True,... }};     
{\[Lambda],    {..., Real -> True,... }};  
\end{verbatim}
\item The scalar component of the singlet superfield receives a VEV after EWSB and splits into scalar and pseudo scalar part:
\begin{equation*}
S = \frac{1}{\sqrt{2}} \left( \phi_S + i \sigma_S + v_S \right)
\end{equation*} 
In the model file this reads as
\begin{verbatim}
{SSing, {{phiS,1/Sqrt[2]},{sigmaS,I/Sqrt[2]},{vS,1/Sqrt[2]}};
\end{verbatim}
\item It is known, that in the CP conserving case, the scalar and pseudo scalar sector does decouple, i.e. the pseudo scalar singlet mixes with the pseudo scalar Higgs and the scalar component with the scalar Higgs:
\begin{verbatim}
{{phid,phiu,phiS},{ZH,hh}},
{{sigmad,sigmau,sigmaS},{ZA,Ah}}
\end{verbatim}
The new eigenstates are CP-even \(h\) (\verb"hh") and CP-Odd \(A_h\) (\verb"Ah") Higgs and the corresponding mixing matrices are called \(Z^H\) (\verb"ZH") and \(Z^A\) (\verb"ZA").  
\item The fermionic component of the singlet fields mixes with the other neutral fermions to five neutralinos
\begin{verbatim}
{{fB, fW0, FHd0, FHu0, conj[FSing]},{ZN, Chi}}
\end{verbatim}
\verb"conj" assigns the complex conjugation. We have to use it here, because we defined the singlet as right handed anti field. 
\item Last step is to define a Dirac spinor from the Weyl spinor for the gauge eigenstates
\begin{verbatim}
dirac[[12]] = {FS, FSing, conj[FSing]};
\end{verbatim}
\end{enumerate}

\paragraph*{Calculating Vertices} After evaluating the model file due to the \verb"Start"-command, we can e.g. calculate the coupling between the scalar Higgs and the d-quarks by
\begin{verbatim}
Vertex[{bar[Fd], Fd, hh}]
\end{verbatim}
and get
\begin{verbatim}
{{bar[Fd[{gt1, ct1}]], Fd[{gt2, ct2}], hh[{gt3}]}, 
{((-I)*conj[ZH[gt3, 2]]*Delta[ct1, ct2]*Delta[gt1, gt2]*
                   Yd[gt2, gt1])/Sqrt[2], PL}, 
{((-I)*conj[ZH[gt3, 2]]*Delta[ct1, ct2]*Delta[gt1, gt2]*
                   Yd[gt1, gt2])/Sqrt[2], PR}}
\end{verbatim}
SARAH writes first a list with the involved particles and inserts the indices, which are used in the result: \verb"gt"X for generation and \verb"ct"X for color indices. The result consists of two entries for left (\verb"PL") and right (\verb"PR") polarization. \\

\paragraph*{Tadpole Equations} SARAH also automatically calculates the minimum conditions for the vacuum: 
\begin{equation*}
\frac{\partial V}{\partial v_i} = 0
\end{equation*} 
We can look at the first tadpole equation  after EWSB  \(\frac{\partial V}{\partial v_d}=0\) via
\begin{verbatim}
TadpoleEquationsEWSB[[1]] 
\end{verbatim}
The output, left hand side of the equation, reads as
\begin{verbatim}
mHd2*vd + (g1^2*vd^3)/8 + (g2^2*vd^3)/8 - (g1^2*vd*vu^2)/8 - 
(g2^2*vd*vu^2)/8 - (3*vS^2*vu*\[Kappa]*\[Lambda])/2 + 
(vd*vS^2*\[Lambda]^2)/2 + (vd*vu^2*\[Lambda]^2)/2 - 
(vS*vu*A[\[Lambda]])/Sqrt[2]
\end{verbatim}

\paragraph*{Ghost Interactions} The neutral Goldstone boson is the first generation of the pseudo scalar Higgs, i.e. \(A_h^1\), and \(\eta_+\) is the corresponding ghost to \(W^+=(W^-)^*\). Thus, if we have defined the gauge fixing terms,
we can also calculate ghost vertices like \(\bar{\eta}_+ \eta_+ A_h^1\) by
\begin{verbatim}
Vertex[{bar[gWmC],gWmC,Ah[{1}]}]
\end{verbatim}
with the following result
\begin{verbatim}
{{bar[gWmC], gWmC, Ah[{1}]}, 
{-(g2*Mass[VWm]*(ZA[1, 2]*Cos[\[Beta]] - ZA[1, 1]*Sin[\[Beta]]))/2, 1}}
\end{verbatim}

\subsubsection{Parametrization of the Pseudo Scalar Sector}
The first implementation of the NMSSM in SARAH runs fine, but it is also common, do define two rotations in the pseudo scalar sector: First isolate the Goldstone by
\begin{equation*}
\left(\begin{array}{c} \sigma_d \\ \sigma_u \\ \sigma_S \end{array} \right) =
\left(\begin{array}{ccc}
\cos\beta & \sin\beta & 0 \\
-\sin\beta & \cos\beta & 0 \\
0 & 0 & 1
\end{array} \right) \left(\begin{array}{c} G \\ \sigma   \\ \sigma_S \end{array} \right)
\end{equation*} 
and rotate afterwards \((\sigma, \sigma_S)\).

Now we want to show how to do this in SARAH:
\begin{enumerate}
\item Define a first, temporary rotation
\begin{verbatim}
Rotation[[1]]={TEMP, ...
{{sigmau, sigmad}, {AhT, ZT}}, ...}
\end{verbatim}
to temporary eigenstates \(A_h^T\) with a mixing matrix \(Z^T\).
\item Parametrize this mixing matrix by
\begin{verbatim}
{ZT, { Real->True,
       Dependence -> {{Cos[\[Beta]], Sin[\[Beta]]},
                      {-Sin[\[Beta]], Cos[\[Beta]]}}
      }}
\end{verbatim} 
\item Make a second rotation to the electroweak eigenstates
\begin{verbatim}
Rotation[[2]]={EWSB, ..., {{AhT,sigmaS}, {Ah, ZA}}, ... }
\end{verbatim}
\item Parametrize the rotation of the second pseudo scalar Higgs and the pseudo scalar singlet by an mixing angle \(\phi\):
\begin{verbatim}
{ZA,  Real -> True,
      Dependence -> {{1, 0,           0},
                     {0, Cos[\[Phi]], -Sin[\[Phi]]},
                     {0, Sin[\[Phi]], Cos[\[Phi]]}} }}
\end{verbatim}
\end{enumerate}

\subsubsection{Adding CP Violation}
So, now we want to drop the assumption about CP conservation in the Higgs sector. This is done just by setting
\begin{verbatim}
{\[Kappa],     {..., Real -> False,... }};     
{\[Lambda],    {..., Real -> False,... }};  
\end{verbatim}
This leads to the fact that the Higgs matrix is no longer reducible and we have to define it as
\begin{verbatim}
{{phid,phiu,phiS,sigmad,sigmau,sigmaS},{ZH,h}}
\end{verbatim}

\paragraph*{Analyzing Mass Matrices} We want to have a look at the \(6 \times 6\) mass matrix of the Higgs sector in SARAH. This mixing was the 4. mixing we defined in the model file, so we can access the corresponding mass matrix just by \verb"MassMatricesFullEWSB[[4]]". The mixing of the scalar down Higgs with the pseudo scalar up Higgs is the (1,5) component of this matrix. Therefore,
\begin{verbatim}
MassMatricesFullEWSB[[4,1,5]]
\end{verbatim}
returns
\begin{verbatim}
  ((-I/2)*vS*A[\[Lambda]])/Sqrt[2]  
+ ((I/2)*vS*conj[A[\[Lambda]]])/Sqrt[2] 
- ((3*I)/4)*vS^2*\[Lambda]*conj[\[Kappa]] 
+ ((3*I)/4)*vS^2*\[Kappa]*conj[\[Lambda]]
\end{verbatim}
This is the expected result: We see that for real values of \(\kappa\) and \(\lambda\) the contributions would cancel and the matrix decouples again as it should be. 

\subsubsection{CKM Basis}
Although we have added CP violation in the Higgs sector, we have neglected so far the fact of flavor violation in the standard quark sector. But it is known that the Yukawa matrices are not diagonal in the gauge eigenstates of the quarks. 
This will lead after EWSB to a rotation of the gauge eigenstates to a new basis, which diagonalizes the Yukawa matrices: 
\begin{eqnarray*}
\hat{d}_L &=& V_d \hat{d}_L^0, \hspace{1cm} \hat{u}_L = V_u \hat{d}_R^0 \\
\hat{d}_R &=& U_d \hat{u}_L^0, \hspace{1cm} \hat{u}_R = U_u \hat{u}_R^0 \\
\end{eqnarray*} 
where we have assigned the gauge eigenstates with a upper index \(0\). We can define diagonal Yukawa \(Y_u, Y_d\) matrices in this new basis by
\begin{eqnarray*}
Y_d = U_d^\dagger Y_d^0 V_d, \hspace{1cm}Y_d = U_u^\dagger Y_d^0 V_d
\end{eqnarray*}
Furthermore, in SUSY theories the soft breaking parameters are rotated in a similar way:
\begin{eqnarray*}
m_q^2 &=& V_d^\dagger {m^0_q}^2 V_d, \hspace{1cm} m_u^2 = U_u^\dagger {m^0_u}^2 U_u, \hspace{1cm} m_d^2 = U_d^\dagger {m^0_d}^2 U_d \\
A_d &=& U_d^\dagger {A_d^0}^T V_d, \hspace{1cm} A_u = U_u^\dagger {A_u^0}^T V_u
\end{eqnarray*} 
After this redefinition the mixing matrices \(V_d,V_u,U_d,U_u\) disappear from the Lagrangian and only the product of \(V_d\) and \(V_u\) stays: 
\begin{equation*}
V_{CKM} = V_u^\dagger V_d
\end{equation*}

Now, we want to implement this steps in our model file in SARAH:
\begin{enumerate}
\item For clearance, we name the gauge eigenstates as \verb"dL0", \verb"uL0", \verb"dR0" and \verb"uR0". Also the unrotated Yukawa matrices get an index \verb"0". The superpotential can now be written as
\begin{verbatim}
{{q0, Hu, u0, Yu0,1}, {q0, Hd, d0, Yd0, -1},...};
\end{verbatim}
\item Afterwards, we rotate the quarks and squarks to the CKM basis
\begin{verbatim}
Rotation[[1]]={SCKM, ...
     {{SdL0}, {SdL, Vd}},     {{SuL0}, {SuL, Vu}},
     {{SdR0}, {SdR, Ud}},     {{SuR0}, {SuR, Uu}},
     {{{FdL0}, {FdR0}}, {{FdL,Vd}, {FdR,Ud}}},
     {{{FuL0}, {FuR0}}, {{FuL,Vu}, {FuR,Uu}}} 
\end{verbatim}
\item Now we define the rotated parameters 
\begin{verbatim}
{Yu0, {Dependence -> 
       sum[i001,1,3]*sum[i002,1,3]*Yu[i001,i002]*Delta[i001, i002]* 
       Vu[i001,index1]*conj[Uu[i002,index2]] }}, 
...
{mq02, {Dependence -> 
        sum[i001,1,3]*sum[i002,1,3]*mq2[i001,i002]*Delta[i001, i002]*
        Vd[i001,index1]*conj[Vd[i002,index2]] }},
...
\end{verbatim}
\item And the CKM matrix as product of the left mixing matrices
\begin{verbatim}
{CKM, {MatrixProduct -> {Vd,Vu}}}
\end{verbatim}
\end{enumerate}

\subsection{Adding bilinear R-Parity Violation}
As next step we want to consider the possibility of bilinear R-Parity violation. This can be done in the following way: We write a superpotential coupling between the gauge singlet, the left handed leptons and the up Higgs. If we now consider the gauge singlet as right neutrino, we get the \(\mu \nu\)SSM \cite{mnSSM1,mnSSM2}. 
\begin{itemize}
\item Define the right handed neutrinos
\begin{verbatim}
Sing[[4]] = {vR, 3, v, 0, 0,  0};
\end{verbatim}
\item Define the superpotential 
\begin{verbatim}
TriW = {...,{l, Hu, v, Yv, 1}, 
            {v ,v, v,\[Kappa],1/3}, 
            {Hu, Hd, v, \[Lambda],-1}};
\end{verbatim}
\item Give the left and right handed neutrinos a VEV
\begin{verbatim}
 {SvL, {vL, 1/Sqrt[2]}, {sigmaL, I/Sqrt[2]},
       {phiL, 1/Sqrt[2]}},
 {SvR, {vR, 1/Sqrt[2]}, {sigmaR, I/Sqrt[2]},
       {phiR, 1/Sqrt[2]}} 
\end{verbatim}
\item Define the mixings in the Higgs sector. The sneutrinos mix with the neutral Higgs and the selectrons with the charged Higgs:
\begin{verbatim}
 {{phiu, phid, phiR, phiL}, {hh, ZH}},
 {{sigmau, sigmad,sigmaR,sigmaL}, {Ah, ZA}},
 {{conj[SHup],SHdm, SeL, SeR},{Hpm,ZP}}
\end{verbatim}
\item Define the mixings in the fermion sector. The neutrinos mix with the neutralinos and the eletrons with the charginos:
\begin{verbatim}
{{fB, fW0, FHd0, FHu0, conj[FvR], FvL}, {L0, ZN}}, 
{{{fWm, FHdm, FeL}, {fWp, FHup, conj[FeR]}}, {{Lm,Um}, {Lp,Up}}}
\end{verbatim}
\end{itemize}

\subsection{Adding an additional Gauge Group}
Let's say, we want now to add an additional \(U(1)\) to the model. This is done in SARAH just by adding a new gauge superfield:
\begin{verbatim}
Gauge[[4]]={X, 1, extra, g4};
\end{verbatim}
The name of the superfield is just \verb"X", so the new vector boson is called \verb"VX" and the new gaugino \verb"fX". The dimension of the group is \verb"1" and we called it \verb"extra" in comparison to \verb"hypercharge" or \verb"color". The parameter for the coupling constant is \verb"g4".\\
As second step we have to define the charge of all chiral under this new group,e.g.
\begin{verbatim}
Doub[[4]] = {Hup, Hu0, 1, Hu,  1/2, 1, 0, -1/2};
\end{verbatim}
This means up Higgs has hypercharge \(\frac{1}{2}\), transforms under \(SU(2)_L\), does not transform under \(SU(3)_C\) and has charge \(-\frac{1}{2}\) with respect to the new gauge group.\\
Obviously, the Higgs couple to the new gaugino \(\tilde{X}\) (\verb"fX" in SARAH). This leads to a further expansion of the neutralino mass matrix by 
\begin{verbatim}
{{fB, fW0, FHd0, FHu0, conj[FvR], FvL, fX}, {L0, ZN}}
\end{verbatim}

But, when we start now the evalution in SARAH, we get this messages:
\begin{verbatim}
WARNING! U(1)^3 (extra) Gauge Anomaly!
WARNING! U(1)*SU(2)^2 (extra x left^2) Gauge Anomaly!
\end{verbatim}
That means, we have not been very careful in choosing the new gauge quantum numbers and should improve this choice. 

\end{appendix}


\begin{thebibliography}{99}


\bibitem{susy1}
J.~Wess and B.~Zumino,
Nucl.\ Phys.\  {\bf B~70} (1974) 39;
P.~Fayet and S.~Ferrara,
Phys.\ Rept.\  {\bf 32} (1977) 249.



\bibitem{susy2}
H.P.~Nilles,
Phys.\ Rept.\  {\bf 110} (1984) 1.

\bibitem{susy3}
H.E.~Haber and G.L.~Kane,
Phys.\ Rept.\  {\bf 117} (1985) 75.


\bibitem{susy4}
E.~Witten,
Nucl.\ Phys.\ {\bf B~188} (1981) 513.

\bibitem{susy5} S.~Dimopoulos, S.~Raby, and F.~Wilczek, 
Phys. Rev.~{\bf D~24} (1981) 1681;
L.E.~Ib\'a\~nez and G.G.~Ross, Phys.~Lett.~{\bf B~105} (1981) 439;
U.~Amaldi, W.~de Boer, and H.~F\"urstenau, Phys.~Lett.~{\bf B~260} 
           (1991) 447;
P.~Langacker and M.~Luo, Phys.~Rev.~{\bf D~44} (1991) 817;
J.~Ellis, S.~Kelley, and D.V.~Nanopoulos, Phys.~Lett.~{\bf B~260} (1991) 161.

\bibitem{susy6}
L.E.~Ib\'a\~nez and G.G.~Ross,
Phys.\ Lett.\ {\bf B~110} (1982) 215.

\bibitem{susy7}
J.R.~Ellis et al.,
Nucl.\ Phys.\ {\bf B~238} (1984) 453.


\bibitem{SARAH}
F.~Staub,
``SARAH,''
arXiv:0806.0538 [hep-ph].

\bibitem{staub2}
F.~Staub, W.~Porod and J.~Niemeyer,
``Strong dark matter constraints on GMSB models,''
arXiv:0907.0530 [hep-ph].

\bibitem{ohl}
T.~Ohl,
``Drawing Feynman diagrams with Latex and Metafont,''
Comput.\ Phys.\ Commun.\  {\bf 90} (1995) 340
[arXiv:hep-ph/9505351].

\bibitem{FormCalc}
T.~Hahn,
``FormCalc 6,''
arXiv:0901.1528 [hep-ph].

\bibitem{micrOmegas}
G.~Belanger, F.~Boudjema, A.~Pukhov and A.~Semenov,
``micrOMEGAs2.0: A program to calculate the relic density of dark matter  in
a generic model,''
Comput.\ Phys.\ Commun.\  {\bf 176} (2007) 367
[arXiv:hep-ph/0607059].

\bibitem{FeynArts}
T.~Hahn,
``Generating Feynman diagrams and amplitudes with FeynArts 3,''
Comput.\ Phys.\ Commun.\  {\bf 140} (2001) 418
[arXiv:hep-ph/0012260].

\bibitem{CHep}
A.~Pukhov {\it et al.},
``CompHEP: A package for evaluation of Feynman diagrams and integration  over
multi-particle phase space. User's manual for version 33,''
arXiv:hep-ph/9908288.

\bibitem{NMSSM}
M.~Maniatis,
``The NMSSM reviewed,''
arXiv:0906.0777 [hep-ph].

\bibitem{mnSSM2}
A.~Bartl, M.~Hirsch, A.~Vicente, S.~Liebler and W.~Porod,
``LHC phenomenology of the $\mu\nu$SSM,''
JHEP {\bf 0905} (2009) 120
[arXiv:0903.3596 [hep-ph]].


\bibitem{mnSSM1}
D.~E.~Lopez-Fogliani and C.~Munoz,
``Proposal for a new minimal supersymmetric standard model,''
Phys.\ Rev.\ Lett.\  {\bf 97} (2006) 041801
[arXiv:hep-ph/0508297].


\end{thebibliography}
\end{document}